# Carrier Concentration Dependencies of Magnetization & Transport in $Ga_{1-x}Mn_xAs_{1-y}Te_y$

M.A. Scarpulla[1,2], K.M. Yu[2], W. Walukiewicz[2], O.D. Dubon[1,2]

[1]Department of Materials Science & Engineering, University of California Berkeley, Berkeley, CA 94720
[2]Lawrence Berkeley National Laboratory, Berkeley, CA 94720

**Abstract.** We have investigated the transport and magnetization characteristics of $Ga_{1-x}Mn_xAs$ intentionally compensated with shallow Te donors. Using ion implantation followed by pulsed-laser melting, we vary the Te compensation and drive the system through a metal-insulator transition (MIT). This MIT is associated with enhanced low-temperature magnetization and an evolution from concave to convex temperature-dependent magnetization.

## INTRODUCTION

$Ga_{1-x}Mn_xAs$ has been established as the prototypical III-V system displaying ferromagnetism mediated by (fairly) delocalized carriers. Despite recent advances in film processing that have allowed the $T_C$ to reach 150 K [1], its dependence on the hole concentration is still poorly understood. Compensation by Mn interstitials, which are both electrically and magnetically active, has complicated magnetic and transport characterization of this system - thus obscuring the underlying physics. Previously, Satoh *et al*. investigated Sn compensation of $Ga_{0.987}Mn_{0.013}As$ films grown by LT-MBE [2]. In this work, we have endeavored to further probe the dependencies of magnetization and charge transport on hole concentration by introducing compensating shallow Te donors in films having much higher Mn concentrations.

## EXPERIMENTAL

Semi-insulating GaAs (001) wafers were implanted with 160 keV $Te^+$ to varying doses followed by a dose of $1.84 \times 10^{16}$ /$cm^2$ 80 keV $Mn^+$. Mn was implanted first to reduce sputtering effects and water cooling was employed to ensure full amorphization. Samples were irradiated in air with single pulses of varying fluence from a KrF excimer laser as we have described previously [3,4]. SIMS measurements reveal that both the Mn and Te concentrations vary with depth and that the peak Mn concentration is roughly x=0.07 for the samples discussed herein. The total amounts of Te and Mn remaining in the films after processing were measured using RBS and PIXE, and we define the overall Te:Mn ratio as γ [4]. We emphasize that no post-irradiation annealing is necessary to achieve high $T_C$s using our process because our films are virtually free of Mn interstitials due to the liquid-phase epitaxial regrowth. All samples were etched in HCl for 15 minutes to remove metallic droplets and Mn oxides from the surface. Magnetization measurements along {110} in-plane directions were carried out using a DC SQUID magnetometer and were shown to be unchanged by the HCl etching. Resistivity measurements were conducted using pressed In contacts in the van der Pauw geometry.

## RESULTS & DISCUSSION

Fig. 1 presents the sheet resistivity of representative $Ga_{1-x}Mn_xAs_{1-y}Te_y$ samples spanning the compensation-induced MIT. For clarification, we emphasize that this MIT is distinct from the MIT commonly observed at low temperatures in $Ga_{1-x}Mn_xAs$. This MIT should be compared to the Anderson localization transition in doped semiconductors, keeping in mind that magnetic interactions may also be significant. The MIT appears to occur in the range of γ~2/3-3/4 for different processing conditions. The γ=0 and 0.64 samples both display metallic behavior typical of $Ga_{1-x}Mn_xAs$ with $x>\sim0.03$. Both samples exhibit a strong resistivity peak near their $T_C$s of 130 and 99 K which is commonly attributed to critical scattering. No distinct peak is present for the insulating $\gamma=0.96$ sample, which displays thermally activated transport between 4.2 and 300 K. However, a drastic change in activation energy occurs near its ferromagnetic $T_C$ of ~68 K.

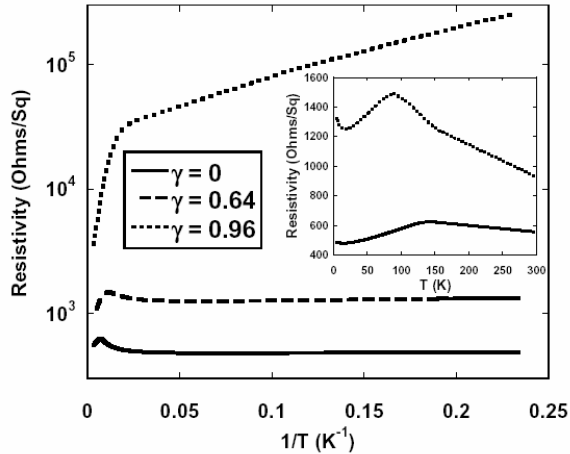

**FIGURE 1.** [main] Resistivity vs. 1/T for $Ga_{1-x}Mn_xAs_{1-y}Te_y$ samples irradiated at 0.2 J/cm$^2$ and having Te:Mn ratios $\gamma=0$, $\gamma=0.64$, and $\gamma=0.96$. [inset] Sheet resistivity vs. temperature for the $\gamma=0$ (solid) and $\gamma=0.64$ (dashed) samples.

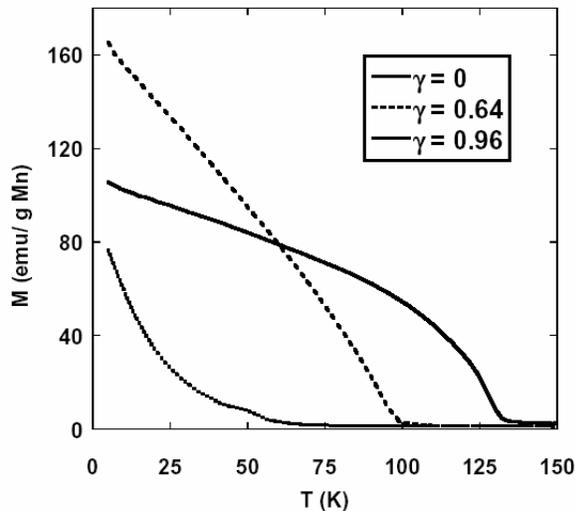

**FIGURE 2.** Magnetization vs. temperature for the set of $Ga_{1-x}Mn_xAs_{1-y}Te_y$ samples presented in Fig. 1.

Fig. 2 presents low-field (50 Oe) measurements of magnetization vs. temperature for the same set of films. The uncompensated $\gamma=0$ sample exhibits a concave dependence, the $\gamma=0.64$ sample near the MIT exhibits a nearly linear dependence, and the strongly-compensated $\gamma=0.96$ sample displays a convex dependence. These changes in magnetization curvature with compensation are qualitatively consistent with recent mean field models [5]. An as-yet unexplained enhancement in low- and high- field magnetizations is present at low temperatures in samples near the MIT.

## CONCLUSIONS

The magnetization and resistivity of Te compensated $Ga_{1-x}Mn_xAs$ films produced using ion implantation and pulsed laser melting were investigated as functions of temperature. A metal insulator transition was identified which delineates metallic films showing concave magnetization curves from insulating films exhibiting thermally activated conduction at all temperatures and convex magnetization curves.

## ACKNOWLEDGMENTS

This work was supported by the Director, Office of Science, Office of Basic Energy Sciences, Division of Materials Sciences and Engineering, of the US Department of Energy under Contract No. DE-AC03-76SF00098. MAS acknowledges support from an NSF Graduate Research Fellowship. ODD acknowledges support from the Hellman Family Fund.

## REFERENCES

1. Ku, K.C., et al., *Appl. Phys. Letters* **82**, 2302-2304 (2003).

2. Satoh, Y., et al., *Physica E* **10**, 196-200 (2001).

3. Scarpulla, M.A., et al., *Appl. Phys. Letters* **82**, 1251-1253 (2003).

4. Scarpulla, M.A., et al., *Physica B* **340-342**, 908-912 (2003).

5. Das Sarma, S., et al., *Phys. Rev. B* **67**, 155201 (2003).